\documentclass[twocolumn,preprintnumbers,amsmath,amssymb]{revtex4}

\usepackage{graphicx}% Include figure files
\usepackage{dcolumn}% Align table columns on decimal point
\usepackage{bm}% bold math
\usepackage{color}

\begin{document}

%\preprint{APS/123-QED}

\title{Nine-Path Quantum Interferometry over 60~km}

\author{Batiste Galm\`es$^{1}$}
\author{Kien Phan-Huy$^{1}$}
\author{Luca Furfaro$^{1}$}
\author{Yanne K. Chembo$^{1,2}$}
\thanks{yanne.chembo@femto-st.fr}
\author{Jean-Marc Merolla$^{1}$}
\thanks{jeanmarc.merolla@univ-fcomte.fr}
\affiliation{$^1$FEMTO-ST Institute, Univ. Bourgogne Franche-Comt\'e, CNRS, Optics Department, \\
                 15B Avenue des Montboucons, 25030 Besan\c{c}on cedex, France.\\
             $^2$GeorgiaTech-CNRS Joint International Laboratory [UMI 2958], Atlanta Mirror Site, \\ 
                 School of Electrical and Computer Engineering, 777 Atlantic Dr NW, Atlanta GA 30332, USA.}

\date{\today}

\begin{abstract}
\textbf{The archetypal quantum interferometry experiment yields an interference pattern that results from 
the indistinguishability of two spatiotemporal paths available to a photon or to a pair of entangled photons.
A fundamental challenge in quantum interferometry is to perform such experiments with a higher number of paths, and 
over large distances. In particular, the distribution of such highly entangled states in long-haul optical fibers 
is one of the core concepts behind quantum information networks.
We demonstrate that using indistinguishable frequency paths instead of spatiotemporal ones allows for robust, high-dimensional quantum interferometry in optical fibers. 
In our system, twin-photons from an Einstein-Podolsky-Rosen (EPR) pair are offered up to $9$~frequency paths after propagation in long-haul optical fibers, and we show that the multi-path quantum interference patterns can be faithfully restored after the photons travel a total distance of up to $60$~km.}
\end{abstract}

\maketitle

%%

%%%%%%%%%%%%%%%%%%%%%%%%%%%%%%%%%%%%%%%%%%%%%%%%%%%%%%%%%%%%%%%%%%%%%%%%%%%%%%%%%%%%%%%%%%%%%%%%%%%%%%%%%%
%%%%%%%%%%%%%%%%%%%%%%%%%%%%%%%%%%%%%%%%%%%%%%%%%%%%%%%%%%%%%%%%%%%%%%%%%%%%%%%%%%%%%%%%%%%%%%%%%%%%%%%%%%
%%%%%%%%%%%%%%%%%%%%%%%%%%%%%%%%%%%%%%%%%%%%%%%%%%%%%%%%%%%%%%%%%%%%%%%%%%%%%%%%%%%%%%%%%%%%%%%%%%%%%%%%%%
\begin{figure}[t]
\centering
\includegraphics[width=7cm]{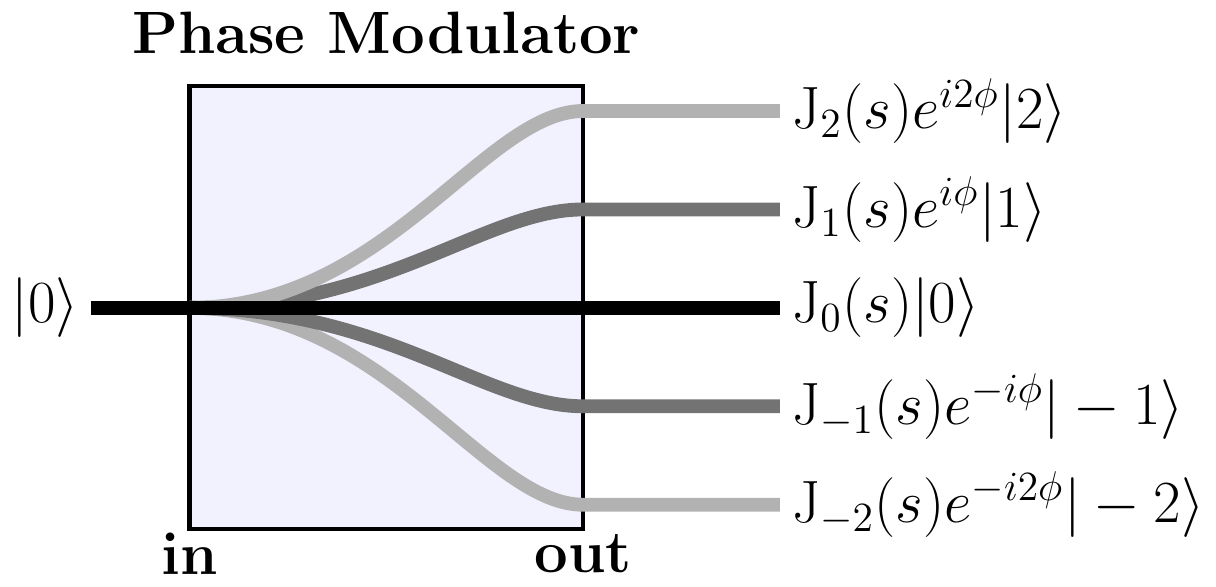}
\caption{Principle of multiple frequency paths using a phase modulator seeded with a photon of frequency $\omega_0$. 
The modulator is driven by a sinusoidal RF signal of amplitude $s V_{\pi}$, frequency $\Omega$ and phase $\phi$. 
This modulation creates the eigenstates $|n \rangle \equiv |\omega_0 + n\Omega \rangle$ with $n \in \mathbb{Z}$, which are new frequency paths available with probability $\mathrm{J}_n^2(s)$ to any incoming photon in the eigenstate $|0 \rangle$.
Only $5$~eigenstates are represented here for clarity, but the entangled photons explore
up to $9$~frequency paths in our experimental transmission system. Along with robustness, this high dimensionality is a distinctive advantage of frequency path quantum interferometry in comparison to spatiotemporal schemes.}
\label{Fig_Quantum_paths_PM}
\end{figure}
%%%%%%%%%%%%%%%%%%%%%%%%%%%%%%%%%%%%%%%%%%%%%%%%%%%%%%%%%%%%%%%%%%%%%%%%%%%%%%%%%%%%%%%%%%%%%%%%%%%%%%%%%%
%%%%%%%%%%%%%%%%%%%%%%%%%%%%%%%%%%%%%%%%%%%%%%%%%%%%%%%%%%%%%%%%%%%%%%%%%%%%%%%%%%%%%%%%%%%%%%%%%%%%%%%%%%
%%%%%%%%%%%%%%%%%%%%%%%%%%%%%%%%%%%%%%%%%%%%%%%%%%%%%%%%%%%%%%%%%%%%%%%%%%%%%%%%%%%%%%%%%%%%%%%%%%%%%%%%%%

%%
Quantum information networks require the manipulation and transportation of entangled photons in long-haul optical fiber networks without destroying their non-classical correlations. 
The relevance and viability of any prospective quantum network also critically depends on a strong hardware overlap with off-the-shelf components, readily available in the technologically mature sector of optical fiber telecommunications.  
Fulfilling these constraints while preserving the non-classical correlations of entangled photons 
when launched in long-haul optical fibers arises as particularly difficult task, and even more so 
if the entanglement is high-dimensional. 
We present here a proof-of-concept quantum interferometry experiment in which 
high-dimensional quantum entanglement is preserved even after the photons have propagated in long-haul optical fibers.

Our system is based on the idea of Franson interferometry~\cite{franson_two-photon_1991}. 
We pump a $3$~cm-long periodically poled lithium niobate (PPLN) waveguide around $775$~nm to generate
entangled photons via spontaneous parametric down conversion (SPDC) around $\lambda_0 = 2 \pi c/\omega_0 = 1550$~nm.
%%%%%%%%%%%%%%%%%%%%%%%%%%%%%%%%%%%%%%%%%%%%%%%%%%%%%%%%%%%%%%%%%%%%%%%%%%%%%%%%%%%%%%%%%%%%%%%%%%%%%%%%%%
%%%%%%%%%%%%%%%%%%%%%%%%%%%%%%%%%%%%%%%%%%%%%%%%%%%%%%%%%%%%%%%%%%%%%%%%%%%%%%%%%%%%%%%%%%%%%%%%%%%%%%%%%%
%%%%%%%%%%%%%%%%%%%%%%%%%%%%%%%%%%%%%%%%%%%%%%%%%%%%%%%%%%%%%%%%%%%%%%%%%%%%%%%%%%%%%%%%%%%%%%%%%%%%%%%%%%
\begin{figure}[t]
\centering
\includegraphics[width=7cm]{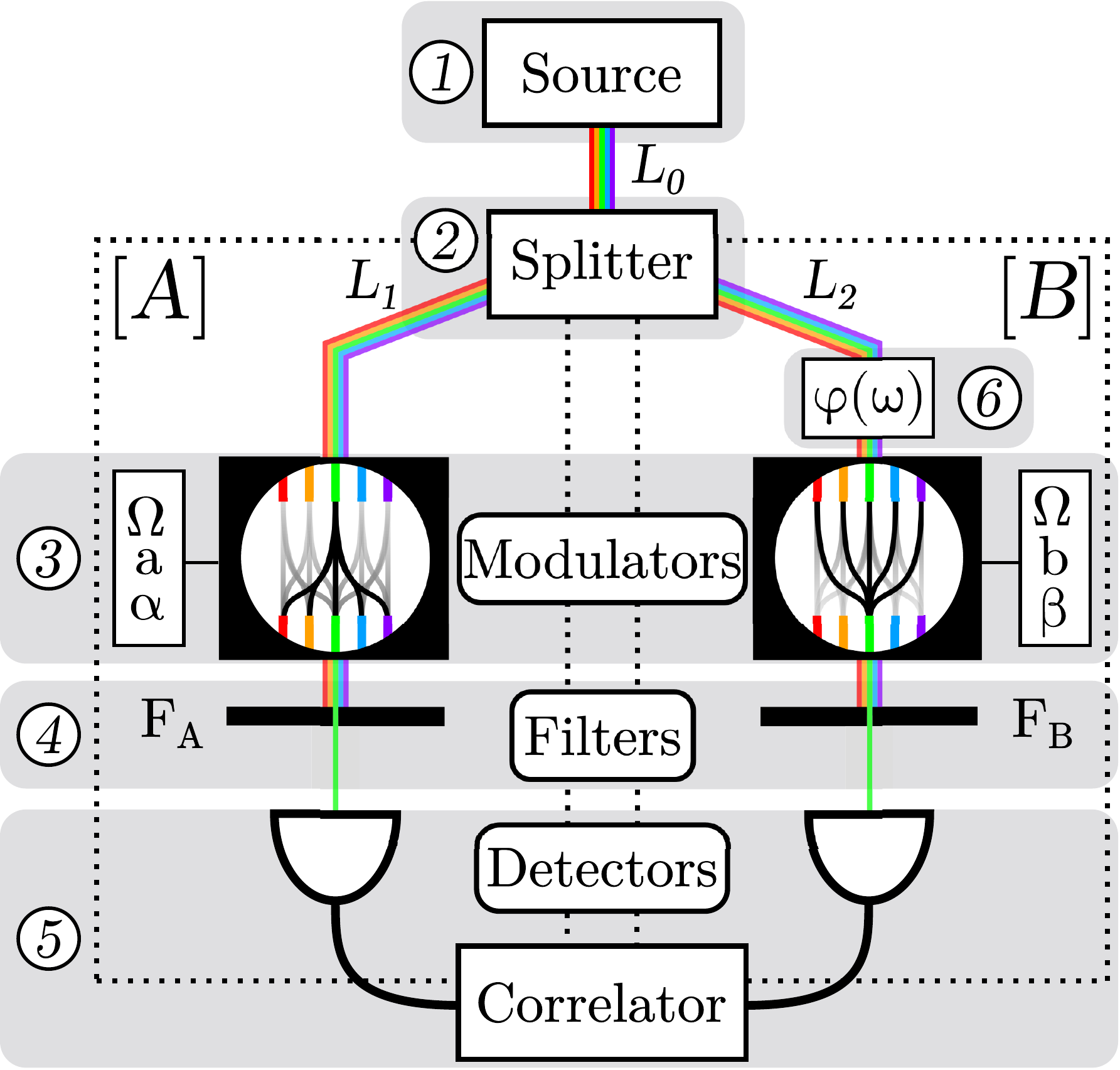}
\caption{Schematic representation of the experimental setup. 
(1)~Broadband twin-photon source. 
(2)~$3$~dB splitter towards arms $[A]$ and $[B]$.
(3)~Phase modulators providing the frequency paths.
(4)~Bragg filters to select the output photon pairs.
(5)~Detection and correlation.  
(6)~Dispersion compensation system.}
\label{Exp_set}
\end{figure}
%%%%%%%%%%%%%%%%%%%%%%%%%%%%%%%%%%%%%%%%%%%%%%%%%%%%%%%%%%%%%%%%%%%%%%%%%%%%%%%%%%%%%%%%%%%%%%%%%%%%%%%%%%
%%%%%%%%%%%%%%%%%%%%%%%%%%%%%%%%%%%%%%%%%%%%%%%%%%%%%%%%%%%%%%%%%%%%%%%%%%%%%%%%%%%%%%%%%%%%%%%%%%%%%%%%%%
%%%%%%%%%%%%%%%%%%%%%%%%%%%%%%%%%%%%%%%%%%%%%%%%%%%%%%%%%%%%%%%%%%%%%%%%%%%%%%%%%%%%%%%%%%%%%%%%%%%%%%%%%%
The spectral density $f(\omega)$ of this source spans over $\sim 2\pi \times 12$~THz around $\omega_0$ ($\sim 100$~nm 
around $\lambda_0$), and the quantum state of the twin-photons is
\begin{eqnarray}
|{\psi}\rangle =  \int_{-\infty}^{+\infty}  d\omega f(\omega) |{\omega_0-\omega}\rangle 
                                                              |{\omega_0+\omega}\rangle \, .
\label{input_state}
\end{eqnarray}
The entangled photons are then launched into a standard SMF-28 fiber of length ${L}_0$ before being separated by a 3~dB coupler in two arms $[A]$ (Alice) and $[B]$ (Bob). 

%%%%%%%%%%%%%%%%%%%%%%%%%%%%%%%%%%%%%%%%%%%%%%%%%%%%%%%%%%%%%%%%%%%%%%%%%%%%%%%%%%%%%%%%%%%%%%%%%%%%%
%%%%%%%%%%%%%%%%%%%%%%%%%%%%%%%%%%%%%%%%%%%%%%%%%%%%%%%%%%%%%%%%%%%%%%%%%%%%%%%%%%%%%%%%%%%%%%%%%%%%%%%%%%
%%%%%%%%%%%%%%%%%%%%%%%%%%%%%%%%%%%%%%%%%%%%%%%%%%%%%%%%%%%%%%%%%%%%%%%%%%%%%%%%%%%%%%%%%%%%%%%%%%%%%%%%%%
\begin{figure*}[t]
\centering
\includegraphics[width=16cm]{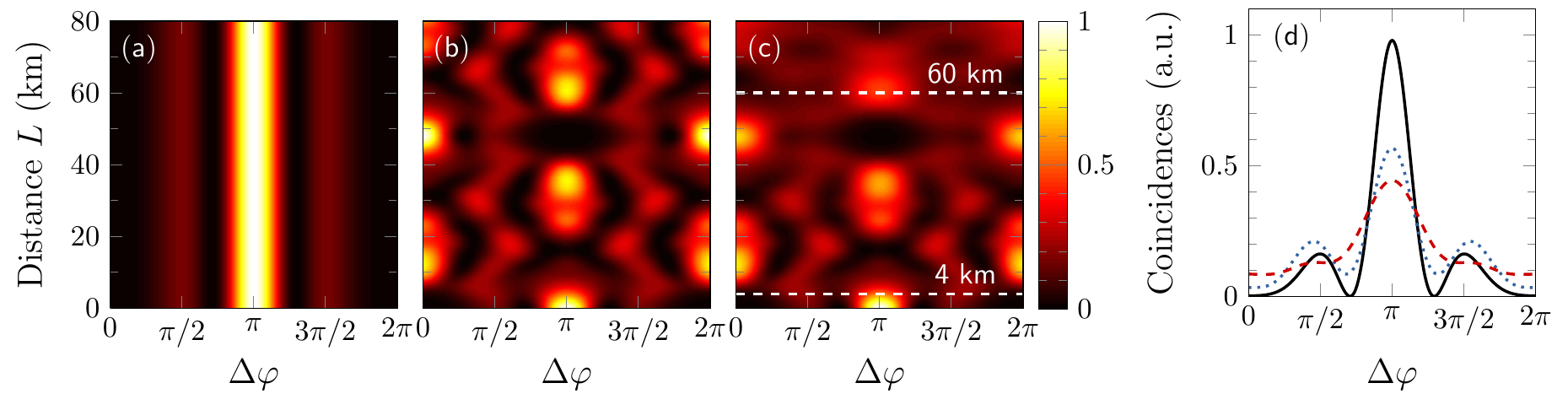}
\caption{Numerical simulation of the interference patterns with respect to the total distance $L$ traveled by the twin-photons, using Eq.~(\ref{Eq_dispersion_prob_int}). 
The parameters are $a=2.8$, $b=2.6$, $\Delta L=0$, and the filters are positioned at $n=0$.  
(a)~Ideal case with no dispersion ($\beta_2 = 0$) and perfectly monochromatic filters  ($\Omega_F=0$).
The interference pattern [which corresponds to the continuous black line in Fig.~\ref{Simulations}(d)] remains unaffected by the propagation.
(b)~Case where dispersion is accounted for ($\beta_2 = -22$~ps$^2$/km) but the filters are still perfectly monochromatic.
The interference pattern evolves periodically, with a period $L_p$ corresponding to length necessary for the accumulated dispersion phase $\beta_2 \Omega^2 L_p/2$ to amount to $2\pi$.
(c)~Realistic case where both the dispersion and the bandwidth of the Bragg filters ($\Omega_F/2\pi=3$~GHz) are accounted for. The interference pattern is irreversibly altered during propagation and impedes long-haul quantum interferometry.
(d)~Quantum interference patterns from Fig.~\ref{Simulations}(c) at 
$L=0$~km (continuous black),  
$L=4$~km (dotted blue), and 
$L=60$~km (dashed red). 
\label{Simulations}}
\end{figure*}
%%%%%%%%%%%%%%%%%%%%%%%%%%%%%%%%%%%%%%%%%%%%%%%%%%%%%%%%%%%%%%%%%%%%%%%%%%%%%%%%%%%%%%%%%%%%%%%%%%%%%%%%%%
%%%%%%%%%%%%%%%%%%%%%%%%%%%%%%%%%%%%%%%%%%%%%%%%%%%%%%%%%%%%%%%%%%%%%%%%%%%%%%%%%%%%%%%%%%%%%%%%%%%%%%%%%%
%%%%%%%%%%%%%%%%%%%%%%%%%%%%%%%%%%%%%%%%%%%%%%%%%%%%%%%%%%%%%%%%%%%%%%%%%%%%%%%%%%%%%%%%%%%%%%%%%%%%%%%%%%

Each arm includes another fiber spool of length ${L}_{A,B}$, which transform the input state after propagation into
\begin{eqnarray}
|{\psi}\rangle = \int_{-\infty}^{+\infty}  d\omega f(\omega) e^{i\Phi(\omega)} 
                                          |{\omega_0 + \omega}\rangle_{A}
                                          |{\omega_0 - \omega}\rangle_{B}
\label{eq_prop}
\end{eqnarray}
where $\Phi(\omega)=\beta(\omega)(L_{0}+L_A)+\beta(-\omega)(L_0+L_{B})$ denotes the phase shift resulting from the propagation in the optical fibers $L_{0,A,B}$, and $\beta(\omega)$ rules the fiber dispersion. 
Whereas Franson interferometry usually involves Mach-Zehnder (MZ) modulators offering two temporal paths~\cite{harris_nonlocal_2008, sensarn_observation_2009, olislager_frequency-bin_2010, olislager_implementing_2012}, our system 
exploits instead $N \gg 2$ frequency paths offered by phase modulators (PM)~\cite{Bloch_99,olislager_creating_2014}. 
The creation of these frequency paths is performed using two phase modulators $\mathrm{PM}_{A,B}$ of half-wave voltage $V_{\pi}$,
modulated at the frequency $\Omega/2\pi = 12.5$~GHz with signals of normalized amplitudes $\{a,b\}=V_{\{a,b\}}/V_{\pi}$ 
and phases $\{\alpha, \beta\}$. The modulators $\mathrm{PM}_{A,B}$ induce the transformations 
\begin{eqnarray}
 |{n}\rangle_{A,B} \rightarrow \sum_{k\in \mathbb{Z}} \mathrm{J}_{k}(\{a,b\}) e^{ik (\{\alpha,\beta\})} 
                               |{n+k}\rangle_{A,B} \, ,
\label{transform_PMA}
\end{eqnarray}
where $|{n}\rangle \equiv |{\omega_0 + n\Omega}\rangle$ are new frequency paths with $n \in \mathbb{Z}$. 
As explained in Fig.~\ref{Fig_Quantum_paths_PM}, every path $|{n}\rangle$ is accessed with probability
$\mathrm{J}_n^2(s)$ where $s \in \{a,b\}$ is the real-valued modulation index of the modulator.
The closure condition $\sum_{n=-\infty}^{+\infty} \mathrm{J}_n^2(s) =1$ is a property of Bessel functions and 
holds for all $s$. However, the probabilities $\mathrm{J}_n^2(s)$ decay rapidly to~$0$ as $n \rightarrow \pm \infty$, and only 
$N \simeq 2(s+1)+1$ frequency paths have a non-negligible probability.  
In our experimental setup, we have $s \simeq 3$ and accordingly, up to $N=9$~paths with $n \in \{-4, \dots, 4\}$ 
can be explored by the twin-photons. 
We use the two Bragg filters $F_{A,B}$ of bandwidth $\Omega_F/2\pi= 3$~GHz, which are respectively centered to 
$\omega_0 + n \Omega$ and $\omega_0 - n \Omega$.
The probability of coincidence measurement by the avalanche photodetectors $D_A$ and $D_B$ is therefore 
\begin{eqnarray}
P_{n} =  \int_{-\Omega_{F}/2}^{+\Omega_{F}/2} \textrm{d} \omega \, g(\omega) \left| \sum_{k\in \mathbb{Z}}
               \mathrm{J}_k(a)   \mathrm{J}_{-k}(b)  e^{i\Phi(k,n,\omega)}  \right|^{2}  ,
\label{Eq_dispersion_prob_int}
\end{eqnarray}
where  $\Phi(k,n,\omega) = k  \left[ \Delta\varphi+\beta_{1} \Omega\Delta L-\beta_{2}(\omega+n\Omega)\Omega  L \right] +  k^{2}\beta_{2}\Omega^{2}L/2$, with $\beta_1$ and $\beta_2$ being respectively the group velocity and the group velocity dispersion in the optical fibers, $L=2L_0+L_A+L_B$ the total distance traveled by the twin-photons, $\Delta L=L_A-L_B$ the distance mismatch, and $g(\omega)$ is spectral density of the source crenelated by the Bragg filters.
The superposition of indistinguishable paths as given by the summation over $k$, and squaring the sum gives rise to quantum interferences. The two-photon interference patterns are obtained by scanning the phase difference 
$\Delta \varphi=\alpha-\beta$. Since the filters have a finite bandwidth $\Omega_F$, the integration of $\omega$ over 
$[-\Omega_F/2,\Omega_F/2]$ results in a continuous superposition of shifted interference patterns leading to a loss of visibility.

The dispersion phase-shift $\Phi(k,n,\omega)$ is such that the $\beta_1$-induced shift is easily canceled with the matching $L_A=L_B$, while the dispersion-induced shift cannot, and therefore has a strong impact on the quantum correlations. 
This phenomenon is illustrated in Fig.~\ref{Simulations} which shows the evolution of numerically simulated interference patterns as the twin-photons propagate in the optical fibers. 
It can be seen that dispersion not only alters the visibility of the fringes, but also disturbs the shape of the interference pattern -- and thus, the quantum correlations. 
However, unlike in the classical case, the problem of dispersion cancellation in the context of quantum interferometry is not conceptually trivial, as dispersion can be canceled either 
\textit{locally}~\cite{steinberg_dispersion_1992,steinberg_dispersion_1992-1,odonnell_observations_2011, cuevas_long-distance_2013, lukens_demonstration_2013, shirai_intensity-interferometric_2014,brendel_measurement_1998, cialdi_nonlocal_2011, baek_nonlocal_2009} or 
\textit{non-locally}~\cite{franson_nonlocal_1992,franson_nonclassical_2009,torres-company_nonlocal_2009,zhong_nonlocal_2013}.
Research is still on-going about the classical or quantum nature of the non-local dispersion 
approach~\cite{shapiro_dispersion_2010, franson_lack_2010}.
We show here that the dispersion can be canceled non-locally, thereby restoring the original interference pattern after the twin-photons are propagated in long-haul optical fibers.

%%%%%%%%%%%%%%%%%%%%%%%%%%%%%%%%%%%%%%%%%%%%%%%%%%%%%%%%%%%%%%%%%%%%%%%%%%%%%%%%%%%%%%%%%%%%%%%%%%%%%%%%%%
%%%%%%%%%%%%%%%%%%%%%%%%%%%%%%%%%%%%%%%%%%%%%%%%%%%%%%%%%%%%%%%%%%%%%%%%%%%%%%%%%%%%%%%%%%%%%%%%%%%%%%%%%%
%%%%%%%%%%%%%%%%%%%%%%%%%%%%%%%%%%%%%%%%%%%%%%%%%%%%%%%%%%%%%%%%%%%%%%%%%%%%%%%%%%%%%%%%%%%%%%%%%%%%%%%%%%
\begin{figure*}[t]
\centering
\includegraphics[width=16cm]{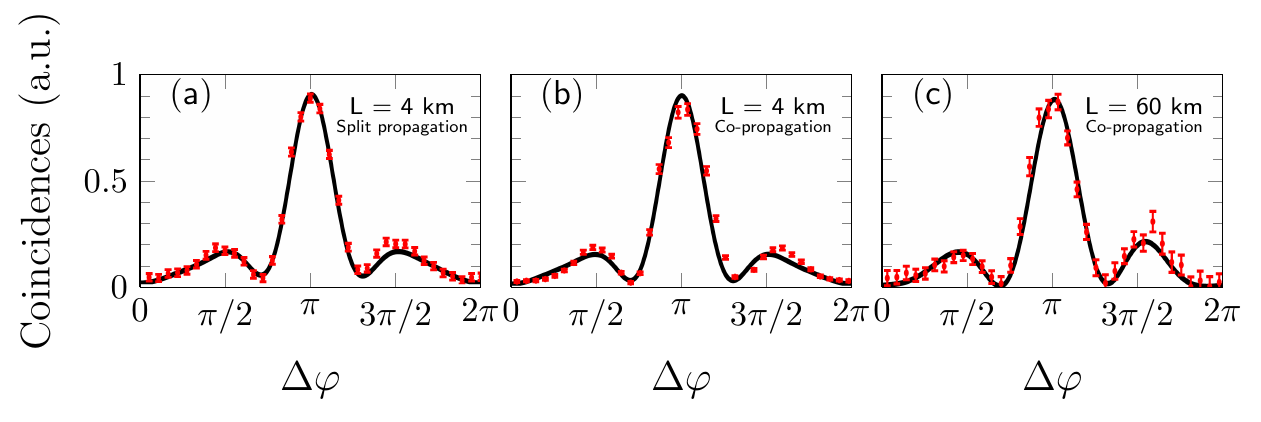}
\caption{
Experimental results for moderate and long distance propagation with our nonlocal dispersion compensation scheme.
The theoretical Bessel-like patterns are the continuous black lines, while the experimental results are indicated with the red symbols.
In all cases, the theoretical Bessel-like interference pattern is restored with high accuracy.
(a)~Interference pattern for $2$~km separated propagation in $L_A$ and $L_B$ ($L=L_A+L_B=4$~km)
(b)~Interference pattern for $2$~km co-propagation in $L_0$ ($L=2L_0=4$~km).
(c)~Interference pattern for $30$~km co-propagation in $L_0$ ($L=2L_0=60$~km).
\label{Fig_Exp_res2} }
\end{figure*}
%%%%%%%%%%%%%%%%%%%%%%%%%%%%%%%%%%%%%%%%%%%%%%%%%%%%%%%%%%%%%%%%%%%%%%%%%%%%%%%%%%%%%%%%%%%%%%%%%%%%%%%%%%
%%%%%%%%%%%%%%%%%%%%%%%%%%%%%%%%%%%%%%%%%%%%%%%%%%%%%%%%%%%%%%%%%%%%%%%%%%%%%%%%%%%%%%%%%%%%%%%%%%%%%%%%%%
%%%%%%%%%%%%%%%%%%%%%%%%%%%%%%%%%%%%%%%%%%%%%%%%%%%%%%%%%%%%%%%%%%%%%%%%%%%%%%%%%%%%%%%%%%%%%%%%%%%%%%%%%%

%%
When $L_A=L_B$, the dispersion shift $\Phi(k,n,\omega)$ depends only the total propagation distance $L=2L_0+L_A+L_B$ traveled by the twin-photons.  
In fact, the dispersion effect is the same whether both photons co-propagate along the fiber ($L_0$), or only along one of the two arms ($L_{A,B}$). This has two major consequences. First, the whole propagation and dispersion can actually be experienced by only one photon and still lead to the same measurements. Second, a negative dispersion applied anywhere cancels the effect. To perform nonlocal dispersion cancellation, a dispersion-compensation module (DCM) is inserted in arm $[A]$, as shown in
Fig.~\ref{Exp_set}. This module introduces a negative phase-shift that compensates to the group velocity dispersion phase shift.
The coincidence probability of Eq.~(\ref{Eq_dispersion_prob_int}) becomes
$P_n \propto  |\textrm{J}_{0}({c})|^{2}$, where $c=\sqrt{a^{2}+b^{2}+2ab\cos(\Delta\varphi)}$, so that
the initial Bessel-like interference pattern is fully recovered~\cite{olislager_implementing_2012}.

The experimental results are presented in Fig.~\ref{Fig_Exp_res2}.
The acquisition of the interference patterns required stable continuous operation for several days.
In particular, each interference pattern corresponds to continuous data acquisition for a duration of $48$~h.
Photon counting was performed using LynX\'ea photon counters from Aurea Technology with onboard time-correlator. 
One of the main challenges comes from the phase-shift $\beta_{1}\Delta L$  that shifts the interference pattern in 
$\Phi(k,n,\omega)$, since $\Delta L$ cannot be strictly set to $0$ when $L_{A,B}$ is km-long. 
Indeed, for large distances, this term is very sensitive to temperature as $1$~K induces a $2\pi$ phase-shift per kilometer.
A simple way to overcome this difficulty is to propagate most of the distance without separating the photons, split them and cancel the whole dispersion afterward. Consequently, $\Delta L$ is then nulled with the best precision possible, and the temperature fluctuations automatically are compensated for being the same for the twin-photon pair. 
The expected results are identical to the case in which the photons are first separated and then propagate most of the distance in two identical fibers $L_A \simeq L_B$. However, this solution may rise questions with regard to the actual separation of the photons and the ``non-locality'' of the experiment. For this reason, two sets of experiments were performed. 

We first let the photons propagate together in the same fiber $L_0=2$~km with negligible propagation in the arms $[A]$ and
$[B]$ (with $L_A=L_B\simeq 0$). The results are then compared to the experiment in which the photons are separated right after the source ($L_0\simeq 0$), while propagating over $L_A=L_B=2$~km.
In both cases, the total distance traveled by the twin photons is $L=4$~km.
A Finisar WaveShaper 4000S with $10$~GHz resolution is used on arm $[A]$ to compensate for the dispersion, and
Figs.~\ref{Fig_Exp_res2}(a) and~(b) show how the Bessel interference pattern is restored for both experiments, in excellent agreement with the simulations. 
This confirms the fact that co-propagation (along $L_0$) or split propagation (along $L_A$ and $L_B$) of the photon pair yield the same results, and as predicted by the theory, the visibility of the interference pattern is also restored.

Since co-propagation or split propagation eventually results in the same outcome, a second set of experiments was performed. Here, in order to show our ability to cancel the dispersion over large distances, both photons travel together within the same fiber of length $L_0=30$~km while $L_A=L_B\simeq 0$, corresponding to a total distance $L=60$~km for the twin-photons. 
The Finisar DCM is replaced by a DCMX from Teraxion with fixed dispersion cancellation for $60$~km. 
The results are shown in Figure~\ref{Fig_Exp_res2}(c), showing again an excellent recovery of the Bessel interference pattern after the long-haul fiber propagation.

In conclusion, we have demonstrated robust, high-dimensional ($N=9$) frequency-path quantum interferometry
in long-haul optical fibers, in which twin-photons travel a distance up to $60$~km.
We expect such systems to play a major role in future quantum information networks, particularly for quantum-key distribution frequency-based systems~\cite{Bloch_99,Merolla_99}, or combined with resonator-based quantum frequency
combs~\cite{Chembo_2016,Kues_2017}. \\

%\textbf{Acknowledgment} 
The authors would like to acknowledge fruitful interactions with Matthieu Bloch.
%\textbf{Funding.} 
This research has been funded by the Labex ACTION.
It has also been funded by the projects CORPS, IQUINS, and PHYFA 
of the \textit{R\'egion Bourgogne Franche-Comt\'e} in France.

%\textbf{Author contributions.} 
%B.~G., K.~P.-H., L.~F. and J.~M.~M. performed the experiments. 
%B.~G. and J.~M.~M. performed the numerical simulations. 
%All the authors discussed the results and analyzed the data.
%B.~G., Y.~K.~C. and J.~M.~M. wrote the article. 
%The research work was supervised by J.~M.~M.
%
%\textbf{Competing financial interests.} 
%The authors declare no competing financial interests.

\end{document}